\documentclass[graybox]{svmult}

\usepackage{mathptmx}       
\usepackage{helvet}         
\usepackage{courier}        
\usepackage{type1cm}        
\usepackage{makeidx}         
\usepackage{graphicx}        
\usepackage{multicol}        
\usepackage[bottom]{footmisc}

\makeindex             

\begin{document}

\title*{Is Life (or at least socio-economic aspects of it) just Spin and Games~?}
\titlerunning{Is Life just Spin and Games~?}
\author{Shakti N. Menon, V. Sasidevan and Sitabhra Sinha}
\institute{Shakti N. Menon \at The Institute of Mathematical Sciences,
CIT Campus, Taramani, Chennai 600113, India.
\and V. Sasidevan \at Department of Physics, Cochin University of Science and Technology,
Cochin 682022, India.
\and Sitabhra Sinha \at The Institute of Mathematical Sciences, 
CIT Campus, Taramani, Chennai 600113, India.
\email{sitabhra@imsc.res.in}
}
%
%
\maketitle

\abstract*{The enterprise of trying to explain different social and economic phenomena using concepts and ideas drawn from physics
has a long history.  
Statistical mechanics, in particular, has been often seen as most likely to provide the means to achieve this, because it provides a lucid and concrete  
framework for describing the collective behavior of systems  
comprising large numbers of interacting entities. Several physicists  
have, in recent years, attempted to use such tools to throw  
light on the mechanisms underlying a plethora of socio-economic  
phenomena. These endeavors have led them to develop a community  
identity - with their academic enterprise being dubbed as  
``econophysics'' by some. However, the emergence of this field has also  
exposed several academic fault-lines. Social scientists often regard  
physics-inspired models, such as those involving spins coupled to each other, as  
over-simplifications of empirical phenomena. At the same time, while 
models of rational agents who strategically make choices based on  
complete information so as to maximize their utility are commonly  
used in economics, many physicists consider them to be caricatures of  
reality. We show here that while these contrasting approaches may seem  
irreconcilable there are in fact many parallels and analogies between  
them. In addition, we suggest that a new formulation of statistical  
mechanics may be necessary to permit a complete mapping of the  
game-theoretic formalism to a statistical physics framework. This may  
indeed turn out to be the most significant contribution of econophysics.}

\abstract{The enterprise of trying to explain different social and economic phenomena using concepts and ideas drawn from physics
has a long history.  
Statistical mechanics, in particular, has been often seen as most likely to provide the means to achieve this, because it provides a lucid and concrete  
framework for describing the collective behavior of systems  
comprising large numbers of interacting entities. Several physicists  
have, in recent years, attempted to use such tools to throw  
light on the mechanisms underlying a plethora of socio-economic  
phenomena. These endeavors have led them to develop a community  
identity - with their academic enterprise being dubbed as  
``econophysics'' by some. However, the emergence of this field has also  
exposed several academic fault-lines. Social scientists often regard  
physics-inspired models, such as those involving spins coupled to each other, as  
over-simplifications of empirical phenomena. At the same time, while 
models of rational agents who strategically make choices based on  
complete information so as to maximize their utility are commonly  
used in economics, many physicists consider them to be caricatures of  
reality. We show here that while these contrasting approaches may seem  
irreconcilable there are in fact many parallels and analogies between  
them. In addition, we suggest that a new formulation of statistical  
mechanics may be necessary to permit a complete mapping of the  
game-theoretic formalism to a statistical physics framework. This may  
indeed turn out to be the most significant contribution of econophysics.}

\section{Introduction}
\label{sec:1}
The physicist Ernest Rutherford is believed to have once distinguished physics from the other sciences, referring to the latter as
merely ``stamp collecting''~\cite{Bernal1939}. 
While Rutherford may have been exceptional in explicitly voicing the traditional arrogance of physicists to other branches
of knowledge, it is true that the spectacular success of physics in explaining 
the natural world has led many physicists to believe that progress has not 
happened in other sciences because
those working in these fields are not trained to examine observed phenomena 
from the perspective of physics.
Intriguingly, practitioners in several branches of knowledge have also occasionally looked at physics as a model to aspire to,
a phenomenon sometimes referred to as ``Physics-envy''. For instance, the science of economics has undergone such a phase, 
particularly in the late nineteenth century, and concepts from classical physics, such as equilibria and their stability, were central to the 
development of the field during this period~\cite{Mirowski1989}. However, this situation gradually changed starting at the beginning of the twentieth century,
curiously just around the time when physics was about to be transformed by the ``quantum revolution'', and economics
took a more formal, mathematical turn. The development of game theory starting from the 1920s and 1930s eventually provided
a new {\em de facto} language for theorizing about economic and social phenomena. However, despite this apparent ``parting of ways''
between economics and physics, there have been several attempts, if somewhat isolated, throughout the previous century to build
bridges between these two fields. In the 1990s, these efforts achieved sufficient traction and a sub-discipline sometimes referred to as
``econophysics'' emerged with the stated aim of explaining economic phenomena using tools from different branches of
physics~\cite{Sinha2010}.  
 
In earlier times, the branch of physics now known as dynamical systems theory had been a rich source of ideas for economists developing their field.
More recently, however, it has been the field of statistical mechanics,
which tries to explain the emergence of systems-level properties at the macro-scale as a result of interactions between its components
at the micro-scale, that has become a key source of concepts and techniques used to quantitatively model
various social and economic phenomena. The central idea underlying this enterprise of developing statistical mechanics-inspired models is that, while the behavior of individuals may be essentially 
unpredictable, the collective behavior of a large population comprising many such individuals interacting with each other may exhibit
characteristic patterns that are amenable to quantitative analysis and explanation, and could possibly even be predicted. 
This may bring to one's mind the fictional discipline of ``psychohistory'', said to have been devised by Hari Seldon of Isaac Asimov's {\em Foundation} series fame~\cite{Asimov1951}, that aimed to predict the large-scale features of future developments 
by discerning statistical patterns inherent in large populations. Asimov, who was trained in chemistry (and was a Professor of Biochemistry at Boston
University), in fact used the analogy of a gas, where the
trajectory of any individual molecule is almost impossible to predict, although the behavior of a macroscopic volume is strictly constrained
by well-understood laws.

A large number of the statistical mechanics-inspired models for explaining economic phenomena appear to use the framework
of interacting spins. This is perhaps not surprising given that spin models provide perhaps the simplest descriptions of
how order can emerge spontaneously out of disorder. An everyday instance of such
a self-organized order-disorder transition is exemplified by the so-called ``effect of a staring crowd''~\cite{Kikoin1978}.
Consider a usual city street where pedestrians walking along the sidewalk are each looking in different arbitrarily chosen directions.
This corresponds to a ``disordered'' situation, where each component is essentially acting independently and
no coordination is observed globally. If however a pedestrian at some point persistently keeps looking at a particular object
in her field of view (which corresponds to a fluctuation event arising through chance),
this action may induce other pedestrians to also do likewise - even though there may actually be nothing remarkable to look
at. Eventually, it may be that the gaze of almost all pedestrians will be aligned with each other and each of them will be staring
into the same point in space that is devoid of any intrinsic interest.
This situation will correspond to the spontaneous emergence of ``order'' through interactions between the components,  i.e., 
as a result of the pedestrians responding almost
unconsciously to each other's actions. It is of course also possible to have everyone stare towards the same point by having
an out-of-the-ordinary event (a ``stimulus'') happen there. In this case, it will be the stimulus extrinsic to the pedestrians - rather than 
interactions between
the individuals - that causes the transition from the disordered to ordered state. 

The simplest of the spin models, the {\em Ising model}, was originally proposed to understand spontaneous magnetization
in ferromagnetic materials below a critical temperature. It assumed the existence of a large number of elementary spins,
each of which could orient in any one of two possible directions (``up'' or ``down'', say). Each spin is coupled to
neighboring spins through exchange interactions, which makes it energetically favorable for neighboring spin pairs to be both
oriented in the same direction. However, when the system is immersed in a finite temperature environment, thermal
fluctuations can provide spins with sufficient energy to override the cost associated with neighboring spins being oppositely aligned.
The spins could also be subject to the influence of an external field that will break the symmetry between the two
orientations and will make one of the directions preferable to the spins. By associating temperature to the degree of
noise or uncertainty among agents, field to any external influence on the agents and exchange coupling between spins
to interaction between individuals in their social milieu, it is easy to see that the Ising model can be employed to 
quantitatively model a variety of social and economic situations involving a large number of interacting individuals. 
Such modeling is particularly relevant when the question of interest involves qualitative changes that occur in the collective
behavior as different system parameters are varied. The nature  of the transition may also be of much interest as external field-driven 
ordering typically manifests as a first-order or discontinuous transition, while transitions orchestrated entirely through interactions between
the components has the characteristics of a second-order or continuous transition. As the latter is often associated with so-called
power laws, it is not unusual that these are often much sought after by physicists modeling social or economic phenomena
(sometimes to the puzzlement of economists).  

The popularity of spin models in the econophysics community has however not percolated to mainstream social scientists, who,
probably justifiably, find such models to be overly simplified descriptions 
of reality. Many economic and social phenomena are
therefore quantitatively described in terms of game theoretic models, where the strategic considerations of individuals, who
rationally choose between alternatives in order to maximize their utilities or payoffs, come to the fore.
However, such approaches have also been criticized as being based upon an idealized view of the capabilities of individual agents and
of the information that they have access to for making decisions. A complete description of aspects of economic life is possibly neither
provided by spin models nor by game-theoretic ones - but being two very different types of caricatures of reality, an attempt to integrate these
two approaches may provide us with a more nuanced understanding of the underlying phenomena. 
With this aim in view, in the two following sections we describe in brief the essential framework of these two approaches that are
used to understand collective behavior in a population of agents. We show that despite differences, there are in fact many parallels
and analogies between spin model-based and game theoretic approaches to describing social phenomena.
We conclude with the suggestion that the statistical mechanics approach used at present may not be completely adequate for describing
strategic interactions between rational agents that is the domain of game theory. This calls for the development of a new formalism that will allow seamless
integration of statistical mechanics with game theory - which will possibly be the most enduring contribution of econophysics to the 
scientific enterprise.
 
\section{Collective decision making by agents: Spins \ldots}
\label{sec:2}
We can motivate a series of models of the dynamics of collective decision making by agents that differ in terms of the
level of details or information resolution that one is willing to consider.
We begin by considering a group of $N$ agents, each of whom are faced with the problem of having to choose between a finite number of
possible options at each time step $t$, where the temporal evolution of the system is assumed to occur over discrete intervals. 
To simplify matters we consider the special case of binary decisions in which the agents, for instance, simply choose between
``yes'' or ``no''. Thus, in the framework of statistical physics, the state of each agent (representing the choice made by it) can be mapped to
an Ising spin variable $S_i = \pm 1$. Just as spin orientations are influenced by the exchange interaction coupling with their neighbors in
the Ising model, agents take decisions that can, in principle, be based on the information regarding the choices made by other agents (with whom they
are directly connected over a social network) in the past - as well as the memory of its own previous choices.
If an agent needs to explicitly identify the specific choice made by each neighbor in order to take a decision, then this
constitutes the most detailed input information scenario. Here, each agent $i$ considers the choices made by its $k_i$ neighbors in 
the social network of which it is a part (if its
own choices also need to be taken into account we may assume that it includes 
itself in its set of neighbors). Furthermore, each agent $i$ has a memory of the choices made by its neighbors in the preceding $m_i$ time steps.
Thus, the agent, upon being presented with a history represented as a $m_i \times k_i$ binary matrix, has to choose between
$-1$ and $+1$. As there are $2^{m_i k_i}$ possible histories that the agent may need to confront, this calls for formulating
an input-output function $f_i$ for the agent that, given a string of $m_i k_i$ bits, can generate the probability that the agent
will make a particular choice, viz., Pr($S_i = +1$) = $f_i (\{\pm 1, \pm 1, \ldots \pm 1\}_{m_i k_i})$ and with
Pr($S_i = -1$)= $1 - $Pr($S_i = +1$).
In other words, the choice of each agent $i$ will be determined by a function whose domain is a $m_i k_i$-dimensional
hypercube and range is the unit interval $[0,1]$. 

The above situation is simplified by assuming that agents do not know the exact identity of the choices made by each of
its neighbors but only have access to the aggregate information 
as to  how many chose a particular option, e.g., $+1$.
A natural extension of this is the scenario where, 
instead of an explicit network, agents are considered to essentially interact with 
the entire group. Such an effectively ``mean-field'' like situation (where pairwise interactions between spins are replaced by
a self-consistent field representing the averaged effect of interactions of a spin with the collective) will arise when, in particular,
an agent's choice is made on the basis of a global observable that is the record of the outcome of choices made by all agents.
For instance, one can model financial markets in this manner, with agents deciding whether to trade or not in an asset based
entirely on its price, a variable that is accessible to all agents and which changes depending on the aggregate choice behavior 
of agents - with price rising if there is a net demand (more agents choose to buy than sell) and falling if the opposite is true
(more agents choose to sell than to buy). Thus, if $N_+$ and $N_-$ are the number of agents choosing $+1$ and $-1$, respectively,
then agents base their decision on their knowledge of the net number of agents who choose one option rather than the other,
i.e., $N_+ - N_- = \sum_i S_i = N M$, with $M$ being the magnetization or average value of spin state in the Ising model.
In this setting, the choice of the $i$th agent having memory (as stated above) is made using information about the value of $M$ 
in the preceding $m_i$ time steps. Therefore, the input-output function specifying the choice behavior of the agents maps
a string of $m$ continuous variables\footnote{We however note that as there only $N$ agents
whose choices need to be summed, the relevant information can be expressed in $\log_2 (N+1)$ bits.} lying in the interval $[-1,1]$ to a probability for choosing a particular
option, viz., Pr($S_i = +1$) = $f_i (M_1, M_2, \ldots, M_m$) where $M_j$ is the value of magnetization $j$ time steps earlier.
One can view several agent-based models that seek to reproduce the stylized features of price movements in financial markets 
as special cases of this framework, including the model proposed by Vikram {\em et al}~\cite{Vikram2011} that exhibits
heavy-tailed distributions for price fluctuations and trading volume which are quantitatively similar to that observed empirically,
as well as volatility clustering and multifractality. 

A further simplification can be achieved upon constraining the function $f_i$ to 
output binary values, so that Pr($S_i = +1$) 
can only be either $0$ or $1$. The set of functional values realized for all possible values of the argument (i.e., all possible histories
that an agent can confront) which defines the {\em strategy} of the agent can, in this case, be written as a binary string of length
$2^{m \log_2 (N+1)} = (N+1)^m$. It is easy to see that the total number of possible distinct strategies is $2^{(N+1)^m}$.
In reality, of course, many of these possible strategies may not make much sense and one would be focusing on the
subset for which $f_i$ has some well-behaved properties such as monotonicity. To simplify the situation even more, 
the granularity of the information on choices made in the past can be reduced~\cite{Sasidevan2018}. In the most extreme case, the information
about the aggregate or net choice of agents at a particular instant can be reduced to a single bit, viz., sign($M_j$) instead 
of $M_j$. This will be the case, for instance, when one only knows whether a particular option was chosen by the majority or not,
and not how many opted for that choice. The number of possible different histories that an agent may confront is only $2^m$
in this situation and thus, the total number of possible strategies is $2^{2^m}$. The well-known {\em Minority Game}~\cite{Moro2004} can be seen
as a special case of this simplified formalism. It is the very anti-thesis of a coordination game with each agent trying to
be contrary to the majority. In other words, each agent is aiming to use those $f_i$ which would ensure $S_i \times$ sign($M$) $= -1$.  

In the detailed input information scenario described above, a Minority Game (MG) like setting will translate into an Ising model 
defined over a network, where connected spin pairs have anti-ferromagnetic interactions with each other. Such a situation
will correspond to a highly frustrated system, where the large number of energy minima would correspond to the 
various possible efficient solutions of the game. However, if the system remains at any particular equilibrium for all time,
this will not be a fair solution as certain individuals will always form the minority and thus get benefits at the expense
of others. A possible resolution that may make it both efficient and fair is to allow for fluctuations that will force
the collective state to move continuously from one minima to another, without settling down into any single one for a very long time.  

An important feature of the MG is the ability of agents to adapt their strategies, i.e., by evaluating at each time step the performance
or payoff obtained by using each of the strategies, the agent can switch between strategies in order to maximize payoff.  One can ask how the
introduction of ``learning'' into the detailed input information scenario will  affect the collective dynamics of the system.
In the classical MG setting, each agent begins by randomly sampling a small number of $f$s (typically $2$) from the set of all possible
input-output functions and then scores each of them based on their performance against the input at each time step, thereafter choosing the
one with the highest score for the next round. In the detailed information setting, we need to take into  account that an agent will need
to consider the interaction strength it has with each of its neighbors in the social  network it is part of. Thus, agents could adapt based
on their performance not just by altering strategy but also by varying the importance 
that they associate with information arriving from their 
different neighbors (quantified in terms of weighted links). Hence, link weight update dynamics could supplement (or even replace) the 
standard strategy scoring mechanism
used by agents to improve their payoffs in this case. For example, an agent may strengthen links with those neighbors whose past choices
have been successful (i.e., they were part of the minority) while weakening links with those who were unsuccessful.
Alternatively, if agent $i$ happened to choose $S_i$ correctly, i.e., so as to have a sign opposite to
that of sign($M$), while its neighbor agent $j$ chose wrongly, learning may lead to the link from $j$ to $i$ 
becoming positive (inducing $j$ to copy the choice made by $i$ in the future) while the link from $i$ to $j$ becomes negative (suggesting
that $i$ will choose the opposite of what $j$ has chosen). 

It may be worth noting in this context that the role of a link weight update rule on collective dynamics has been investigated
in the context of spin models earlier, although in the different context of coordination where agents prefer to make similar choices 
as their neighbors~\cite{Singh2014}. Using a learning rule that is motivated by the Hebbian weight update dynamics that is often used
to train artificial recurrent neural network models, it has been seen that depending on the rate at which link weights adapt (relative
to the spin state update timescale) and the degree of noise in the system, one could have an extremely high diversity in the time required to 
converge to {\em structural balance} 
(corresponding to spins spontaneously segregating into two  clusters, such that within each cluster all interactions are ferromagnetic and
all interactions between spins belonging to different clusters are anti-ferromagnetic) from an initially frustrated system.
It is intriguing to speculate as to what will be observed if instead the learning dynamics tries to make the spins mis-align with
their neighbors, which would be closer to the situation of MG.

\section{Collective decision making by agents: \ldots and Games}
\label{sec:3}
We now shift our focus from the relatively simpler spin-model inspired descriptions of collective behavior of agents to
those that explicitly incorporate strategic considerations in the decision-making of agents. Not surprisingly, this often
involves using ideas from game theory. Developed by John von Neumann in the early part of the $20$th century, the 
mathematical theory of games provides a rigorous framework to describe decision-making by ``rational'' agents.

It appears intuitive that the states of binary Ising-like spins can be mapped to the different choices of agents when they
are only allowed to opt between two possible actions. We will call these two options available to each agent as action A
and action B, respectively (e.g., in the case of the game Prisoners'  Dilemma, these will correspond to ``cooperation''
and ``defection'', respectively). However, unlike in spin models, in the case of games it is difficult to see in general that
the choices of actions by agents are somehow reducing an energy function describing the global state of the system.
This is because instead of trying to maximize the total payoff for the entire population of agents, each agent (corresponding
to a ``spin'') is only trying to maximize its own expected payoff - sometimes at the cost of others. Possibly the only exception
is the class of the Potential Games wherein one can, in principle, express the desire of every agent to alter their action 
using a global function,  viz., the ``potential'' function for the entire system.

Let us take a somewhat more detailed look into the analogy. For a spin-model, one can write down the effective time-evolution
behavior for each spin from the energy function as the laws of physics dictate that at each time step the spins will try to
adopt the orientation that will allow the system as a whole to travel ``downhill'' along the landscape defined by the
energy function $$E = - \sum_{ij} J_{ij} S_i S_j + h \sum_i S_i.$$ Here, $J_{ij}$ refers  to the strength of interaction between
spins $i$ and $j$, the summation $\sum_{ij}$ is performed over neighboring spin pairs and $h$ refers to an external field.
In the absence of any thermal fluctuations (i.e., at zero temperature), it is easy to see that the state of each spin
will be updated according to $$S_i (t+1) = {\rm sign} (\sum_j J_{ij} S_j + h).$$
For the case of a symmetric 2-person game, the total utility resulting from the choice of actions made by a group of agents
whose collective behavior can be decomposed into independent dyadic interactions, will  be given by $$U = R f_{AA} + P f_{BB} +  (S+T) f_{AB}.$$ 
Here $R$ and $P$ refer to the payoffs obtained by two agents when both choose A or both choose B, respectively, while if one chooses
A and the other chooses B, the former will receive $S$ while the latter will receive $T$. The variables $f_{AA}$,  $f_{BB}$ and $f_{AB}$
refer to the fraction of agent pairs who both choose A, or both choose B, or where one chooses A while the other chooses B, respectively.
On the other hand, for an individual agent the payoff is expressed as
$$U_i = \sum_j p_i p_j R + p_i (1-p_j) S + (1-p_i) p_j T + (1-p_i)(1-p_j) P,$$ where $p_i, p_j$ refer to the probabilities of agents $i$ and
$j$, respectively, to  choose action A. As an agent $i$ can only alter its own strategy by varying $p_i$, it will evaluate $\partial U_i/\partial p_i$
and increment or decrement $p_i$ so as to maximize $U_i$, eventually reaching an equilibrium.

\begin{figure}[tbp]
\includegraphics[width=.99\linewidth]{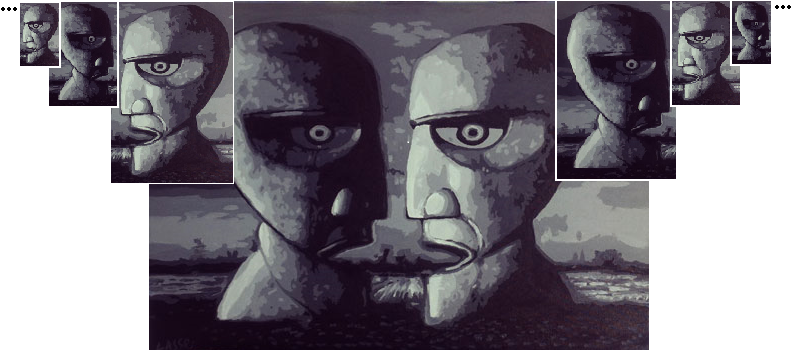}
\caption{A schematic diagram illustrating the infinite regress of theories of mind 
(viz., "she thinks that I think that she thinks that I think that \ldots'')
that two opponents use to guess the action that the other will choose. 
Figure adapted from a drawing of the cover of the {\em Division Bell} music 
album of Pink Floyd designed by
Storm Thorgerson based on illustrations by Keith Breeden.}
\label{fig:2}       
\end{figure}

Different solution concepts will be manifested according to the different ways an agent can model the possible strategy $p_j$ used
by its opponent $j$
(which of course is unknown to the agent $j$). Thus, in order to solve the above equation the agent $i$ actually replaces the variable  
$p_j$ by its assumption $\hat{p_j}$ about that strategy.
In the conventional Nash solution framework, the agent is agnostic about its opponent's strategy so that even $\hat{p_j}$ is an unknown.
To physicists, this approach may sound similar to that of a maximum entropy formalism where the solution is obtained with the least
amount of prior knowledge about the situation at hand. 
However, advances in cognitive science and attempts to develop artificial intelligence capable of semi-human performance in various
tasks have made us aware that human subjects rarely approach a situation where they have to anticipate their opponent's move
with a complete `blank slate' (so to say). Even if the opponent is an individual who the subject is encountering for the first time,
she is likely to employ a {\em theory of mind} to try to guess the strategy of the opponent. Thus, for example, a goalie facing a
penalty kick will make a decision as to whether to jump to the left or the right as soon as the kick is taken (human response time
is too slow for it to make sense for the goalie to  wait until she actually sees which direction the ball is kicked) by trying to simulate
within her mind the thought process of the player taking the kick. In turn, the player
taking the penalty kick is also attempting to guess whether the goalie
is more likely to jump towards the left or the right, and will, so to say,
try to ``get inside the mind'' of the goalie. Each player is, of course,
aware that the other player is trying to figure out what she is thinking and will
take this into account in their theory of mind of the opponent. 
A little
reflection will make it apparent that this process will ultimately lead to an 
infinite regress where each individual is modeling the thought process of 
the opponent simulating her own thought process, to figure out what the
opponent might be thinking, and so on and so forth (Fig.~\ref{fig:2}).
  
The co-action solution framework~\cite{Sasidevan2015, Sasidevan2016} solves
the problem of how agents decide their strategy while taking into account
the strategic considerations of their opponent
by assuming that if both agents 
are rational, then regardless of what exact steps are used by each to arrive
at the solution, they
will eventually converge to the same strategy. Thus, in 
this framework,  $\hat{p_j} = p_i$. This results in solutions that often 
differ drastically from those obtained in the Nash framework. For example,
let us consider the case of the 2-person {\em Prisoners' Dilemma} (PD), a well-known
instance of a {\em social dilemma}. Here, the action chosen by each of the agents 
in order to maximize their individual payoffs paradoxically results in
both of them ending up with a much inferior outcome than that would have been
obtained with an alternative set of choices. In PD, each agent has the choice
of either cooperation (C: action A) or defection (D: action B) and the payoffs
for each possible pair of actions chosen by the two (viz., CC, DD, CD or DC) have the hierarchical
relation $T >  R >  P >  S$. The value of the payoff $T$ is said to
quantify the temptation of an agent for unilateral defection, while $R$ is
the reward for mutual cooperation, $P$ is the penalty paid when both agents
choose defection and $S$ is the so-called ``sucker's payoff'' obtained
by the agent whose decision to cooperate has been met with defection by its opponent.
Other symmetric 2-person games can be defined upon altering the hierarchy among the values of the different
payoffs.  Thus, $T > R > S > P$ characterizes a game referred to as {\em Chicken} (alternatively referred to as
{\em Hawk-Dove} or {\em Snowdrift}) that has been used extensively to model phenomena ranging from nuclear sabre-rattling 
between nations (with the prospect of mutually assured destruction) to evolutionary biology.
Another frequently studied game called {\em Stag Hunt}, which is used to
analyze social situations that require agents to coordinate their actions in order
to achieve maximum payoff, is obtained when $R > T \geq P > S$.

In the Nash framework, the only solution to a one-shot PD (i.e., when the game is played only once) is for both agents to choose defection. 
As is easily seen, they therefore end up with $P$,  whereas if they had both cooperated they would have received $R$ which is a higher
payoff. This represents the dilemma illustrated by the game, namely that choosing to act in a way which appears to be optimal for the individual
may actually yield a sub-optimal result for both players. Indeed,
when human subjects are asked to play this game with each other, they are often seen to
instinctively choose cooperation over defection. While this may be explained by assuming irrationality on the part of
the human players, it is worth noting that the apparently naive behavior on the part of the players actually allows them to
obtain a higher payoff than they would have received had they been strictly ``rational'' in the Nash sense.
In fact, the rather myopic interpretation of rationality in the Nash perspective may be indicative of more fundamental
issues. As has been pointed out in Ref.~\cite{Sasidevan2015}, there is
a contradiction between the two assumptions underlying the Nash solution,  viz., (i) the players are aware that they are both
equally rational and (ii) that each agent is capable of {\em unilateral deviation}, i.e., to choose an action that 
is independent of what its opponent does. The co-action framework resolves this by noting that
if a player knows that the other is just as rational as her, she will take this into
account and thus realize that both will eventually use the same strategy (if not the same action, as in the case of a mixed strategy).
Therefore, cooperation is much more likely in the solution of PD in the co-action framework, which is in line with empirical observations.

A much richer set of possibilities emerges when one allows the game to be played repeatedly between the same set of agents. 
In this iterative version of PD (IPD), mutual defection is no longer the only solution even in the Nash framework, because
agents need to now take into account the history of prior interactions with their opponents. Thus, direct reciprocity between
agents where, for example, an act of cooperation by an agent in a particular round is matched by a reciprocating act of cooperation by its
opponent in the next round, can help in maintaining cooperation in the face of the ever-present temptation towards unilateral defection.
Indeed, folk theorems indicate that mutual cooperation is a possible equilibrium solution of the infinitely repeated IPD.
Multiple reciprocal strategies, such as ``tit-for-tat'' and ``win-stay, lose-shift''  have been devised and their performance
tested in computer tournaments for PD. Intriguingly, it has been shown that when 
repeated interactions are allowed between rational agents, the co-action solution is for agents to adopt a Pavlov strategy.
In this, an agent sticks to its previous choice if it has been able to achieve a sufficiently high payoff but alters the choice if it
receives a low payoff, which allows robust cooperation to emerge and maintain itself~\cite{Sasidevan2016}. 

Moving beyond dyadic interactions to general $N$-person games, the analysis of
situations where an agent simultaneously interacts with multiple neighbors 
can become a formidable task, especially with increasing number of agents.
Thus, one may need to simplify the problem considerably in order to investigate
collective dynamics of a group of rational agents having strategic interactions
with each other. One possible approach - which deviates somewhat from the
strictly rational nature of the agents - invokes the concept of 
{\em bounded rationality}. Here, the ability of an agent to find the optimal
strategy that will maximize its payoff is
constrained by its cognitive capabilities and/or the nature of the
information it has access to.
A notable example of such an approach is the model proposed by Martin Nowak and 
Robert May~\cite{Nowak1992},
where a large number of agents, arranged on a lattice,
simultaneously engage in PD
with all their neighbors in an iterative fashion. As in the conventional 
2-player iterated PD, each agent may choose to either cooperate or defect at each 
round, but with the difference that the agents nominate a single 
action that it uses in its interactions with each of its neighbors. At the end 
of each round, agents accumulate the total payoff received from each 
interaction and compares it with those of its neighbors. It then copies the action
of the neighbor having the highest payoff to use in the next round. 
Note that each 
agent only has access to information regarding the decisions of agents 
in a local region, viz. its topological neighborhood, and hence the 
nature of the collective dynamics is 
intrinsically dependent on the structure of the underlying connection 
network. Nowak and May demonstrated that the model can 
sustain a non-zero fraction of cooperating agents, even after a very 
large number of rounds. In other words, limiting interactions to an agent's
network neighborhood may 
allow cooperation to remain a viable outcome - a concept that has been referred
to as {\em network reciprocity}.

The model described above has been extremely influential, particularly
in the physics community, where it has motivated
a large number of studies that have built upon the basic framework provided
by Nowak and May.
Beyond the implications for how cooperation can be sustained in a population
of selfish individuals, 
these studies have revealed tantalizing links between game theory and 
statistical physics. For instance, by considering the distinct 
collective dynamical regimes as phases, one may describe the switching 
between these regimes in terms of non-equilibrium phase transitions. 
The non-equilibrium nature is manifest from the breakdown of detailed balance
(where the transition rate from one state to another is exactly matched
by that of the reverse process) because of the existence of absorbing
states. These states are defined by cessation of further evolution once
they are reached by the system and correspond to either all agents being 
cooperators or all being defectors. The system cannot escape these states
as agents can only copy actions that are still extant in the population.

While Nowak and May had considered a deterministic updating procedure
(viz., the `imitate the best' rule described above), there have been 
several variants that have incorporated the effect of 
uncertainty in an agent's decision-making process. One of the most 
commonly used approaches is to allow each agent $i$ to choose a 
neighbor $j$ at random and copy its action with a probability given 
by the Fermi distribution function:
\[
\Pi_{i\rightarrow j}=\frac{1}{1+\exp(-(\pi_{j}-\pi_{i})/K)}\,,
\]
where $\pi_{i}$ and $\pi_{j}$ are, respectively, the total payoffs 
received by agents $i$ and $j$ in the previous round, and $K$ is the 
effective temperature or noise in the decision-making process~\cite{Szabo1998}. The 
utility of this function is that it allows one to smoothly interpolate 
between a deterministic situation in the limit $K\rightarrow 0$ (viz.,
agent $i$ will copy agent $j$ if $\pi_{j}>\pi_{i}$) and a completely 
random situation in the limit $K\rightarrow\infty$ (viz., agent $i$ 
will effectively toss a coin to decide whether to copy agent $j$).
Implementing this scheme in a population of agents whose interactions are
governed by different connection topologies allows us to investigate 
the spectrum of collective dynamical states that arise, and the transitions
between them that take place upon varying system parameters~\cite{Menon2018}.

%
\begin{figure}[tbp]
\includegraphics[width=.99\linewidth]{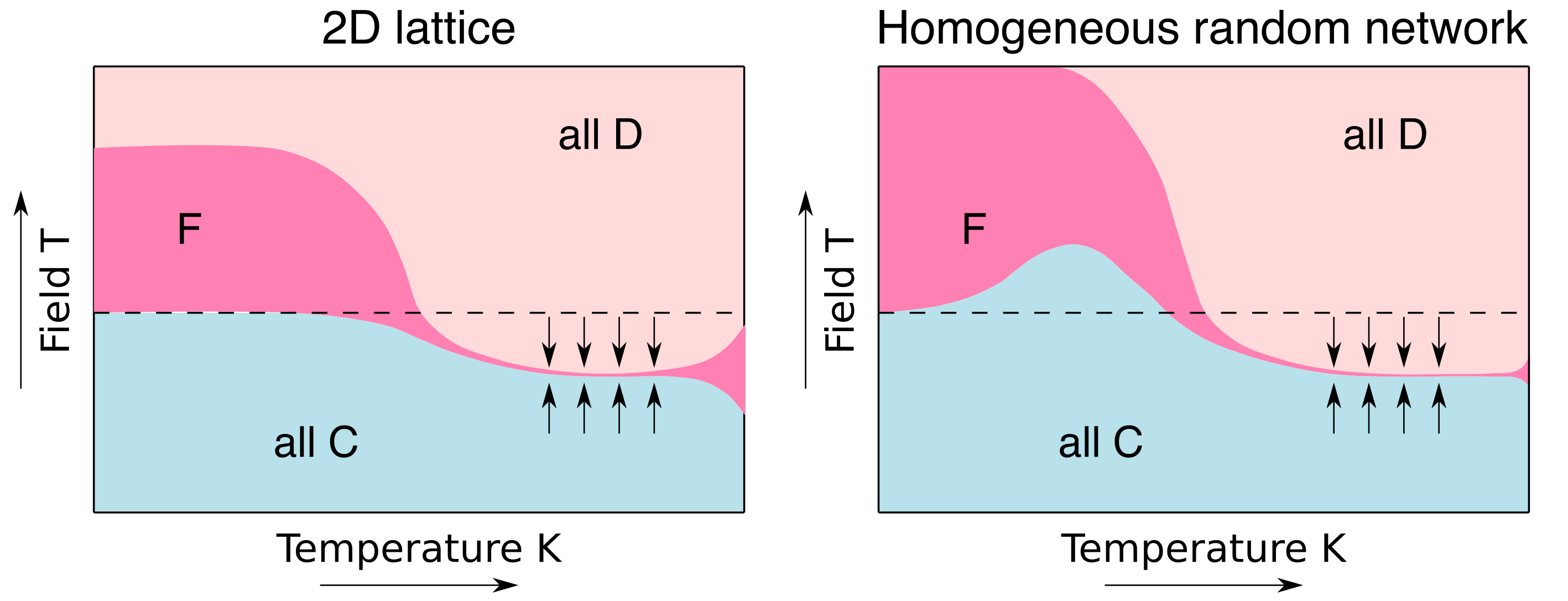}
\caption{Schematic parameter space diagrams illustrating the dependence on the contact network  
structure of the collective dynamics of a system of agents that  
synchronously evolve their states (representing actions) through strategic interactions with  
their neighbors. Each agent in the system adopts one of two possible actions  
at each round, viz. cooperate or defect, and receives an accumulated  
payoff based on each of their neighbor’s choice of action. The agents  
update their action at every round by choosing a neighbor at random  
and copying their action with a probability that is given by a Fermi  
function, where the level of temperature (noise) is controlled by the  
parameter $K$. The broken horizontal line in both panels corresponds to the case  
where the temptation $T$ (payoff for choosing defection when other agent has chosen cooperation) is equal to the reward $R$ for  
mutual cooperation. Hence the region above the line corresponds to the case  
where agents play the Prisoner’s Dilemma game, while that below  
corresponds to the case where they play the Stag Hunt game. 
Note that the temptation $T$ can be viewed as a field, in analogy to 
spin systems, as its value biases an agent's preference for which 
action to choose.
The three  
regimes displayed in each case correspond to situations where the  
system converges to a state where all the agents cooperate (``all  
C''), all agents choose defection (``all D'') or the states of the agents  
fluctuate over time (``F''). We note that the region corresponding to fluctuations appears to  
comprise two large segments connected by a narrow strip. However, the nature of the  
collective behavior is qualitatively different in the two segments,  
as the dynamics observed for large $K$ can be understood as arising  
due to extremely long transience as a result of noise. The left panel  
displays the regimes obtained when agents are placed on a two-dimensional lattice,  
where each agent has $8$ neighbors, while the right panel displays  
the situation where agents are placed on a homogeneous random network  
where all nodes have $8$ neighbors. The difference in the collective  
dynamics between the two scenarios is most noticeable at intermediate  
values of $K$, where the system can converge to an all C state even in  
the Prisoner’s Dilemma regime.}
\label{fig:1}       
\end{figure}
Fig.~\ref{fig:1} shows the different collective states of the system that occur at various regions of the ($K,T$) parameter space.
It is tempting to compare this with the phase diagrams obtained by varying the temperature and external field in spin systems.
First,  the state of an agent, i.e., the action chosen by it at a particular time instant, can be mapped to a spin orientation - e.g.,
if the $i$th agent chooses cooperation, then the corresponding spin state can be designated $S_i = +1$, whereas $S_i = -1$
implies that the agent has chosen defection. 
Typically, there is symmetry between the two orientations $\{-1,+1\}$ that a spin can adopt. However, in games such as PD
one of the actions may be preferable to another under all circumstances (e.g., unconditional defection or $p=0$ is the dominant
strategy in PD). This implies the existence of an effective external field, whose magnitude is linearly related to the ratio of the
temptation for defection and reward for cooperation payoffs, viz., $1-(T/R)$, that results in one of the action choices being more
likely to be adopted by an agent than another. We also have noise in the state update dynamics of the agents as, for a finite
value of $K$, an agent stochastically decides whether to adopt the action of a randomly selected neighbor who has a higher 
total payoff than it. This is not unlike the situation where spins can 
sometimes switch to energetically unfavorable orientations because of thermal 
fluctuations,
when the system is in a finite temperature environment.

Analogous to ordered states in spin systems (corresponding to the spins being aligned), we
have the collective states all C (all agents choose to cooperate) or all D (all agents have chosen defection),
and similar to a disordered state we observe that the collective dynamics of agents can converge to a fluctuating state F
where agents keep switching between cooperation and defection. Just as in spin systems, the phases are
distinguished by using an order parameter, namely, magnetization per spin $m= \sum_i S_i /N \in [-1,1]$, we can define
an analogous quantity $2 f_C - 1$, which is a function of the key observable for the system of agents, viz., the fraction
of agents who are cooperating at any given time $f_C$. As for $m$, the value of this quantity is bounded between $-1$
(all D) and $+1$ (all C), with the F state yielding values close to $0$ provided sufficient averaging is done
over time. 

Note that despite this analogy between the parameters (viz., temperature/noise and field/payoff bias) governing the 
collective dynamics of spin systems and that of a population of agents that exhibit 
strategic interactions with each other,
there are in fact significant differences between the two. As is manifest from Fig.~\ref{fig:1}, an increase in the noise $K$
does not quite have the same meaning as raising the temperature in spin systems. Unlike the latter situation,
agents do not flip from cooperation to defection with equal probability as the temperature/noise increases.
Instead, with equal probability agents either adopt the action chosen by a randomly selected neighbor or stick  
to their current action state. Not surprisingly, this implies that all C and all D states will be stable (for different
values of the field $T$, the payoff value corresponding to temptation for unilateral defection, relative to the reward for
mutual cooperation)  even when $K$ diverges.

In addition, even in the absence of noise (i.e., at $K=0$) we observe that agents can keep switching between different
actions. In other words, unlike the situation in spin systems at zero temperature, the system will keep evolving dynamically.
When an agent determines that a randomly selected neighbor has higher total payoff than it, the agent will 
switch to the action chosen by its neighbor deterministically. Therefore, if there is a coexistence of cooperation and
defection states there will  be switching between these two actions - thereby ensuring the existence
of the fluctuating state at $K=0$.   

Spin systems are also characterized by coarsening dynamics, wherein spins of 
similar orientation coalesce over time to
form domains. Existence of such domains in a spin system, whereby spins of 
opposite orientations can coexist even in the ordered
phase, mean that even at low temperatures the global magnetization of a sufficiently large systems can yield quite small values.
This happens not because of the absence of order, as is obvious, but because of 
coexistence of ordered regions that happen
to be oppositely aligned. At the boundary of two such domains, 
the existence of spin pairs that are oppositely aligned means
that there is a energy cost which increases with the perimeter of the boundary. Thus, energy minimization will result in the
boundaries becoming smoother over time and the shape of the domains eventually stabilize.

Agents on lattices or networks
will also exhibit the spontaneous formation of domains or clusters of interacting agents who have chosen the same action.
Indeed, in order to maintain cooperation in the system for any length of time (in the presence of defectors), the cooperators
will have to form clusters. Within these clusters agents receive a sufficiently
high payoff from cooperating neighbors to prevent them from switching
to defection, despite the potential for being exploited by any neighbor that
chooses to defect.
However, the collective dynamics leads to a form of ``anti-coarsening". This is because agents choosing defection would like to
be surrounded by as many cooperating agents as possible in order to maximize their payoff, so that the boundary between
groups of cooperators and defectors will tend to develop kinks and corners over time, instead of becoming smoother as in
the case of spins. Furthermore, as the cooperators would tend to prefer as few defectors as possible at neighboring positions,
we would observe ceaseless flux in the shape of the domain boundaries unless the system eventually converges to
any one of the two absorbing states, all C or all D.

As already mentioned earlier, the mechanism of agents copying the action of neighbors who are more successful than them - although
helping to simplify the dynamics - is somewhat dis-satisfactory as the agents are now no longer strictly rational. For instance, if the
collective dynamics results in the system converging to the all C absorbing state, all agents will always cooperate with each other  
from that time onwards, as there is no agent left to copy the defection action from. Yet, in a one-shot PD game, defection is always the dominant
strategy as will be realized by any agent who is being ``rational'' and works out the implications of its action in light of the payoff 
matrix (instead of blindly copying its neighbor). Of course, in the iterated PD, it is no longer true that unconditional defection is the best
strategy~\cite{Axelrod1984}. Nevertheless, an all C state is highly unstable 
as it provides a lucrative target for
agents who choose to defect, knowing that they will reap an extremely high payoff 
at the expense of the cooperators. One possible way to prevent global cooperation
from being an absorbing state
in the modeling framework described above is to introduce a mutation probability.
This will allow agents to spontaneously switch
to a particular action with a low probability, independent of whether any of their more successful neighbors is using it or not.
This will ensure that even if a population has reached an all C state, it need not remain there always. 

A more innovative
approach that re-introduces the essential rationality of agents in the context of studying the collective dynamics of a large number of agents
interacting over a social network has been introduced in Sharma {\em et al}~\cite{Sharma2019}. Although formulated in the specific
context of agents making rational decisions as to whether to get vaccinated (based on information about the incidence of a disease and
knowledge of how many neighbors have already gotten vaccinated), the framework can be generally applied to understand many
possible situations in which a large number of agents make strategic decisions through
interactions with other agents. In this approach, each agent plays a symmetric 
2-person game with its ``virtual self'', rather than with any of its neighbors,
in order to decide its action. The interaction with neighbors is introduced by making specific entries in the payoff
matrix that an agent uses for its decision process into functions of the number of its neighbors who have chosen a particular action.
Thus, in the context of vaccination, if all its neighbors have already chosen to vaccinate themselves, an agent is already protected
from disease and is most likely to choose not to get vaccinated (thereby avoiding any real or imagined cost associated with
vaccination, e.g., perceived side-effects). As the neighborhood of each agent is different (in general) when considering either
a lattice or a network, this means that each agent is playing a distinct game. Not only will the games played by each other
differ quantitatively (i.e., in terms of the payoffs of the game) but also qualitatively. Thus, for instance, one agent may be playing what
is in effect PD while another may be playing Chicken. Initial explorations suggest that such spatio-temporal variation of strategies
may give rise to a rich variety of collective dynamical phenomena, which have implications for problems as diverse as designing
voluntary vaccination programs so as to have maximum penetration in a population and predicting voter turnout in elections.
 
\section{In lieu of a Conclusion}
\label{sec:4}
The brief presentation in this chapter of several approaches towards understanding the collective dynamics of a population of
interacting agents, by using both physics-inspired 
spin models and game theoretic models of rational individuals making  
strategic decisions, has hopefully made it clear that there are clear parallels and analogies between the two
frameworks. Although both are at best caricatures of reality, albeit of different types, comparing and contrasting 
between the results generated by both of these approaches should help us understand better how and why large groups
or crowds behave in certain ways. While physicists may harbor the hope of revolutionizing the understanding of
society through the use of simple models of self-organizing phenomena, it may also be that the contribution may be
the other way around. In general, for a group of rational agents, unlike the case in spin models, there appears to 
be no single global function (such as energy) whose minimization leads to the collective states. Thus, it appears that the traditional 
tools of statistical mechanics maybe inadequate for describing
situations where the same collective state may have different utilities for each agent. For instance, in PD, agent $1$ choosing C while
agent $2$ choosing  D, may be the best of all possible  outcomes for $2$ - but it is the worst of all  possible outcomes for agent $1$.
Therefore, while agent $2$ may be desirous of nudging the system to such an outcome, agent $1$ maybe as vehemently trying
to push the system away from such a state. How then would one proceed to model the collective activity of such systems using
the present tools of statistical mechanics~? It does appear that we may need to have a new formulation of statistical mechanics
that applies to the situation outlined above. Thus, it may well turn out that the lasting significance of econophysics will be
in not what it does for economics, but rather in the new, innovative types of physical theories, particularly in statistical physics, that it may spawn.

\begin{acknowledgement}
We thank our collaborator Anupama Sharma, whose joint work with us forms the basis of several ideas discussed above, and 
Deepak Dhar, whose
insightful comments had first gotten us intrigued about the relation between strategic games and statistical physics. The research reported
here has been funded in part by the Department of Atomic Energy, Government of India through the grant for Center of Excellence in Complex Systems
and Data Science at IMSc Chennai.  
\end{acknowledgement}

\end{document}